Eur. Phys. J. C (2020) 80:867
https://doi.org/10.1140/epjc/s10052-020-8418-4THE EUROPEAN
PHYSICAL JOURNAL C

Regular Article - Experimental Physics

# Characterization of water-based liquid scintillator for Cherenkov and scintillation separation

**J. Caravaca**[1,2,a] , **B. J. Land**[1,2], **M. Yeh**[3], **G. D. Orebi Gann**[1,2]

[1] University of California, Berkeley, CA 94720-7300, USA
[2] Lawrence Berkeley National Laboratory, Berkeley, CA 94720-8153, USA
[3] Brookhaven National Laboratory, Upton, NY 11973-500, USAReceived: 1 June 2020 / Accepted: 29 August 2020
© The Author(s) 2020**Abstract** This paper presents measurements of the scintillation light yield and time profile for a number of concentrations of water-based liquid scintillator, formulated from linear alkylbenzene (LAB) and 2,5-diphenyloxazole (PPO). We find that the scintillation light yield is linear with the concentration of liquid scintillator in water between 1 and 10% with a slope of $127.9 \pm 17.0$ ph/MeV/concentration and an intercept value of $108.3 \pm 51.0$ ph/MeV, the latter being illustrative of non-linearities with concentration at values less than 1%. This is larger than expected from a simple extrapolation of the pure liquid scintillator light yield. The measured time profiles are consistently faster than that of pure liquid scintillator, with rise times less than 250 ps and prompt decay constants in the range of 2.1–2.85 ns. Additionally, the separation between Cherenkov and scintillation light is quantified using cosmic muons in the CHESS experiment for each formulation, demonstrating an improvement in separation at the centimeter scale. Finally, we briefly discuss the prospects for large-scale detectors.## 1 Introduction

Liquid-based optical detectors have been used to great success in particle physics, in particular for neutrino physics and rare event searches (see Sec. 35.3.1 of [1] for a complete overview). A broad community is pursuing the idea of a hybrid optical detector, that can leverage both Cherenkov and scintillation signals simultaneously [2–4]. Such a detector could achieve good energy resolution while also having sensitivity to particle direction. The ratio of Cherenkov to scintillation light would also offer a powerful handle for particle and event classification, while a clean identification of the Cherenkov cone would be ideal for long-baseline neutrino physics. A detector with these capabilities would have favorable signal to background ratio across a broad spectrum of physics topics [5–8].

Given the much higher light yield for scintillation over Cherenkov, the detection of a clean Cherenkov signal from pure scintillator is very challenging. Nevertheless, it has been successfully demonstrated in the CHESS experiment [9] and other cases [10] using a technique that exploits the differences in emission time profiles and topology.

The pursuit of water-based liquid scintillator (WbLS) [11], a mixture of liquid scintillator (LS) in water, is one possible avenue to allow optical detectors to reach these requirements. First, it provides a scintillator with an absorption length close to water, increasing the photon detection efficiency in large scintillator detectors; second, it provides a better energy resolution than that of pure Cherenkov detectors; and third it may enhance the identification of Cherenkov light in scintillator detectors via a reduced scintillation yield. The exact formulation can be optimized to address particular physics goals, including modifying the choice and quantity of fluor to affect both the scintillation yield and time profile.

The goal of this paper is to characterize the scintillation light yield and time profile of several WbLS formulations in order to define a Monte Carlo (MC) model that can be used in the future to predict performance of WbLS in large detectors. We study 1%, 5% and 10% formulations of LS in water, using both $\beta$ and $\gamma$ sources, in all cases using linear alkylbenzene (LAB) as the base LS, with 2,5-diphenyloxazole (PPO) as a fluor. This effort is complementary to [12], where the characterization of WbLS is performed with a pulsed X-ray beam source and no Cherenkov radiation is produced. We also use vertical-going cosmic muons in order to demonstrate detection of Cherenkov rings in WbLS and to further validate our MC model. This manuscript is structured in the following way: details of the MC model of the apparatus and the WbLS behaviour are presented in Sect. 2; the WbLS

[a] e-mail: jcaravaca@berkeley.edu (corresponding author)Published online: 20 September 2020Springer

scintillation light yield and time profile measurements are described in Sects. 3.1 and 3.2, respectively; evaluation of the Cherenkov and scintillation separation for cosmic muons and a comparison to model prediction using the CHESS experiment is shown in Sect. 4; and we conclude with Sect. 5 by holding a qualitative discussion of the performance of WbLS in large-scale detectors.

## 2 Monte Carlo model

A detailed MC simulation of photon creation, propagation and detection is used in this analysis, as included in the GEANT4-based [13] simulation package RAT-PAC [14]. RAT-PAC uses the shielding physics list v2.1, which is the one recommended for low count nuclear physics experiments, one of the main applications targeted by this analysis. Our detector simulation includes the full detector geometry, a 3D photomultiplier (PMT) model, optical properties for all detector media, and a model of data acquisition that is described in detail in [4]. It uses the GLG4Scint model [14] to simulate the scintillation light emission, photon absorption and photon reemission, the Rayleigh scattering process developed by the SNO+ collaboration [15], and the default GEANT4 model for Cherenkov photon production. We refer to [4] for further details on any aspect of the MC simulation.

The inputs to the model from measurements presented in this paper are the absolute, intrinsic scintillation light yield and the scintillation emission time profile. Other parameters are implemented from external measurements or estimations, as described in the remainder of this section. This includes the scintillation emission spectrum, refractive index, the attenuation length, Rayleigh scattering length, and the reemission probability.

### 2.1 Emission spectrum

We use the spectra reported in [12] for each WbLS concentration. The different spectra have been measured to be very close to that of pure LABPPO (2g/L) [12]. This is justified by the fact that the emission process is dominated by the radiative transition in the scintillator; addition of non-scintillation components does not affect the emission spectrum. Furthermore, the emission spectrum of LABPPO for PPO concentrations above 2g/L is very similar to that of the 2g/L [10,12].

### 2.2 Refractive index estimation

In order to estimate the refractive index for WbLS, $n$, Newton's formula for the refractive index of liquid mixtures [16] is used:

$$n = \sqrt{\phi_{LS} n_{LS}^2 + (1 - \phi_{LS}) n_W^2}, \quad (1)$$

where $\phi_{LS}$ is the volume fraction of LABPPO in WbLS, and $n_{LS}$ and $n_W$ correspond to the measured refractive indexes for LABPPO [15] and water [17], respectively. Due to the large water concentration, the WbLS refractive index is very similar to that of pure water.

### 2.3 Absorption and scattering lengths

The absorption length, $\lambda$, of WbLS input to the MC model depends on the molarity $c$ of each of the components and, thus, on the concentration of liquid scintillator in water. It is computed as:

$$\lambda = \left(\varepsilon_{lab} c_{lab} + \varepsilon_{ppo} c_{ppo} + \varepsilon_{water} c_{water}\right)^{-1}, \quad (2)$$

where $\varepsilon_{lab}$, $\varepsilon_{ppo}$ and $\varepsilon_{water}$ are the molar absorption coefficients of LAB, PPO [15] and water ([18] for wavelengths over 380 nm and [19] for wavelengths below 380 nm).

In the same fashion, the Rayleigh scattering length, $\lambda^s$, for WbLS is estimated as:

$$\lambda^s = \left(\varepsilon_{lab}^s c_{lab} + \varepsilon_{water}^s c_{water}\right)^{-1}, \quad (3)$$

where $\varepsilon_{lab}^s$ and $\varepsilon_{water}^s$ are the molar scattering coefficients for LAB [15] and water, respectively. It was noted that the addition of PPO does not change $\lambda^s$, and thus it is avoided in Eq. (3). As in the previous case, the estimated values of these properties for WbLS are very close to those of pure water.

This method might overestimate the attenuation lengths, in particular the component due to scattering, given the complex chemical structure and composition of WbLS. This has negligible impact on the results presented here, due to the small (centimeter-) scale of the apparatus. The impact will be more significant for extrapolations of these results to large detectors. A planned long-arm measurement of absorption and scattering lengths is underway.

### 2.4 Scintillation reemission

A photon absorbed by the scintillator volume has a non-zero probability of being reemitted. This reemission process becomes important at low wavelengths where the absorption is strong, shifting photons to longer wavelengths where the detection probability is higher due to a smaller photon absorption and a greater PMT quantum efficiency. The probability $p_i^{reem}$ of a component $i$ in the WbLS absorbing a photon of frequency $\omega$ is determined as the contribution of the given component to the total absorption coefficient, $\alpha$:

$$p_i^{reem}(\omega) = \phi_i \alpha_i(\omega) / \alpha(\omega), \quad (4)$$

where the variables have been previously defined. After a photon is absorbed, it can be reemitted with a 59% probability for LAB and an 80% probability for PPO [15], following the same primary emission spectrum.

## 3 Methods and measurements

Measurements of the scintillation light yield and time profile, using two distinct bench-top setups, are described in the following section.

### 3.1 Light yield measurement

To determine the absolute scintillation light yield of WbLS (number of photons produced per unit of energy deposited), we measure the number of photoelectrons (NPE) detected by a PMT located in front of a WbLS sample excited with a known radioactive source. We fit our MC model (see Sect. 2), which is previously calibrated, to the NPE distribution, by tuning the intrinsic light yield parameter in the model. Given that we simulate the relevant geometry, optical properties of the materials, Cherenkov production, PMT response and data acquisition system, this technique takes into account the effects introduced by those factors. In this way, the extracted light yield corresponds to an absolute property of the medium, rather than being defined relative to scintillator standards. Details of the apparatus, calibration, analysis methods and results are given below.

#### 3.1.1 Description of the apparatus

A minimal configuration is used to measure the absolute light yield of the WbLS mixtures. A 10 mm × 10 mm × 40 mm, 1-mm thick, high-performance quartz cuvette filled with WbLS is located 10 cm from the front face of a 10-inch R7081 Hamamatsu PMT (see Fig. 1a). The PMT is fixed to a black aluminum structure for mechanical stability to reduce possible geometry-related systematic uncertainties. A $^{90}$Sr beta source [20] is located close to the back of the cuvette so that the betas penetrate the WbLS and the produced photons are collected by the 10-inch PMT. The acquisition is triggered off a 1-inch, cubic Hamamatsu PMT optically coupled to the bottom of the cuvette facing upwards. The portion of the trigger PMT front face which is not coupled to the cuvette is covered with a black mask in order to avoid reflections on the otherwise exposed PMT glass. The set formed by the cuvette, the trigger PMT and the source is held by a 3D-printed black shelf, also attached to the aluminum structure and kept at a fixed distance from the R7081 PMT. Waveforms from both PMTs are digitized by a CAEN V1742.

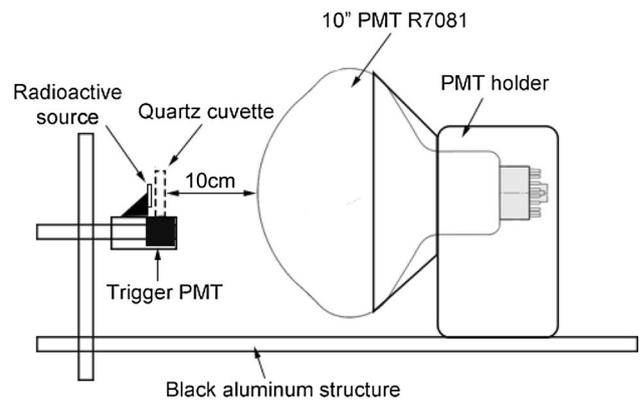

**(a)** Light yield configuration

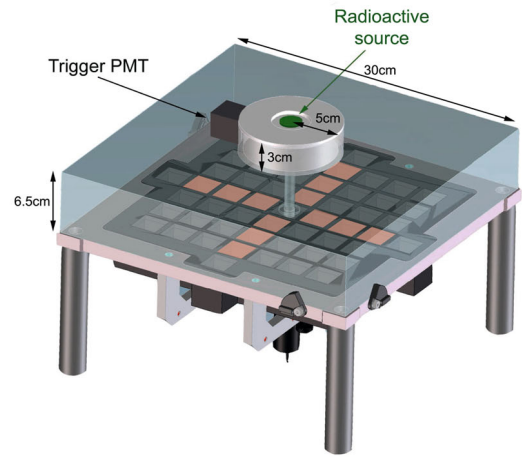

**(b)** Time profile configuration

**Fig. 1** Experimental configurations used in this analysis. Top: scintillation light yield setup; bottom: scintillation time profile configuration. The four scintillator panels that surround the time profile setup on two sides and the bottom are not shown in this figure and we refer to [4] for technical drawings of their layout. See Sects. 3.1.1 and 3.2.1 for further details

#### 3.1.2 Calibration and analysis method

The PMT gain is measured using a dataset with a deionized water target and the PMT located far from the cuvette, in order to ensure single-photoelectron events. The gain is measured by fitting a Gaussian model to the single-PE PMT charge distribution, as described in [9], resulting in $(158.0 \pm 1.0)$ V × ns. The NPE is defined as the total charge collected in the PMT, corrected by the measured PMT gain.

The PMT collection efficiency (CE) is measured using the nominal configuration, with the cuvette filled with deionized water. NPE distributions are compared to the ones obtained using the MC and we fit for a CE parameter, defined as an additional weight to the modeled PMT quantum efficiency. A CE of $87 \pm 8\%$ is obtained. This value is used as the nominal value in the light yield measurements.

**Table 1** Measured absolute light yield parameters. Statistical and systematic uncertainties are included in the quoted numbers

| WbLS | Y [photons/MeV] |
| --- | --- |
| 1% | $234 \pm 30$ |
| 5% | $770 \pm 72$ |
| 10% | $1357 \pm 125$ |
| LABPPO 2g/L | $11{,}076 \pm 1004$ |

For each WbLS sample, a dataset of 100,000 events is collected. The absolute light yield parameter, $Y$, is defined in the MC model as the number of photons $n_\gamma$ emitted per unit of quenched deposited energy $E_q$ given by Birk's law [21], so the total number of photons produced is:

$$n_\gamma = Y \times E_q. \quad (5)$$

We obtain $Y$ by performing a binned maximum likelihood fit to the observed NPE distributions and minimizing $-2 \log \mathcal{L}$.

### 3.1.3 Results

The results for the light yield measurement are shown in Table 1 along with the NPE distributions in Fig. 2. The result for pure LABPPO is included for completeness, and comparison to previous results, with which it is in good agreement. According to our simulation, the region below 8 PE is populated by betas that do not reach the WbLS volume and only produce Cherenkov light in the quartz. Thus, the fit is constrained to the region above 8 PEs. The observed discrepancies in the low PE region are not considered a cause of systematic uncertainty on this measurement given that it is due to events that do not generate scintillation light.

The measured light yield as a function of the LS concentration is shown in Fig. 3. The LABPPO 2g/L case is also shown for completeness and corresponds to the point at 100% concentration. The measured light yield for LABPPO is in agreement with that provided by independent measurements [15], which gives confidence in our MC model and methods. The three WbLS points show a linear behavior with a slope of $127.9 \pm 17.0$ ph/MeV/%LS and an intercept value of $108.3 \pm 51.0$ ph/MeV. Given that the Cherenkov contribution is taken into account by using the full MC model, the non-zero intercept indicates that the light yield is not linear with the LS content at very low concentrations (below 1%). The measured light yield is noticeably larger than that expected from a naive linear extrapolation of the LABPPO 2g/L light yield. The considered sources of systematic uncertainty affecting the light yield measurement are the PMT gain and the PMT CE, whose uncertainties are 0.6% and 9%, respectively, as evaluated from the measurements in Sect. 3.1.2.

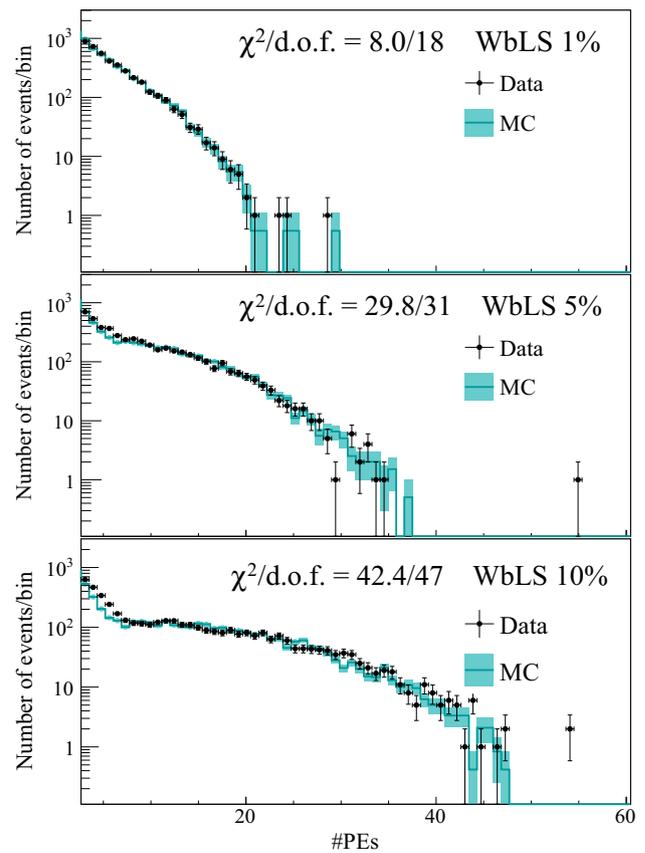

**Fig. 2** Distribution of the number of photoelectrons detected for the three different WbLS concentrations, compared to the result from the absolute light yield fit. Black crosses correspond to data and blue boxes to the best-fit MC. Only statistical uncertainties are shown. The shape corresponds to the characteristic beta spectrum of the $^{90}$Sr source [20]

We ran further tests to ensure our measurements were stable and no other systematic effects were present. Datasets for all the liquid mixtures, including water and LABPPO, were retaken by refilling the cuvette and re-coupling it to the trigger PMT in order to test for stability. No significant changes were observed. In addition, in order to test the performance of our MC model, datasets using different gamma emitters ($^{60}$Co, $^{137}$Cs, $^{22}$Na) in pure LABPPO 2g/L were collected and compared to the MC. The MC prediction using the measured $Y$ was compatible with the data in the three cases.

### 3.2 Time profile measurement

The scintillation time profile of each of the WbLS mixtures produced by excitation with a radioactive source is characterized by obtaining the time profile of the single photoelectrons (SPE) detected by a PMT array and fitting them using our MC model (see Sect. 2). By using a complete MC simulation and by calibrating our detector's SPE response and cable delays, we can account for effects on the time profile due to detector geometry, PMT pulse shape, multi-PE hits or the digitiza-tion process. Thus, the intrinsic time profile of the scintillation emission is extracted, unfolding any of those potential systematic effects. Description of the apparatus, calibration, methods and results can be found below.

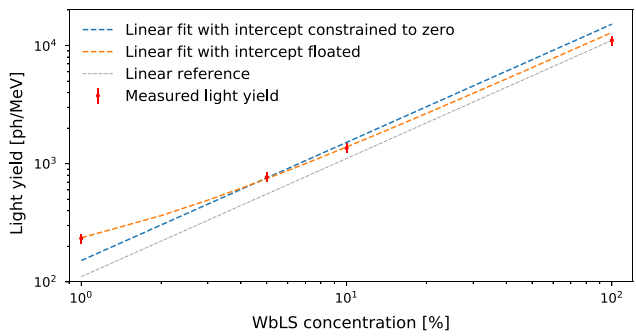

**Fig. 3** Measured absolute scintillation light yield as a function of concentration (red dots, displaying the total uncertainty), including pure LAB-PPO 2g/L (100% concentration). Two linear fits to the three WbLS points are shown, one with the intercept constrained to zero (blue) and another one freely floating the intercept (orange). For reference, a model linear with the pure LABPPO 2/L light yield and with zero intercept is shown (gray)

### 3.2.1 Description of the apparatus

A WbLS sample deployed in a cylindrical acrylic container (referred as the vessel) is excited with betas from a $^{90}$Sr source located on top (see Fig. 1b). Light is collected by an array of 12 1-inch PMTs (Hamamatsu H11934-200) and their waveforms are recorded by a digitizer (CAEN V1742). An optically coupled acrylic block is placed between the PMT array and the vessel to act as a light guide and avoid total internal reflections in the vessel. A PMT (referred as the trigger PMT) is optically coupled to the side of the vessel and provides the data acquisition trigger. The setup features 4 scintillator panels that provide $4\pi$ coverage to the vessel against through-going cosmic rays.

### 3.2.2 Calibration and analysis method

We collect a million triggered events using the $^{90}$Sr source. To veto cosmic ray events we reject those events in which any of the scintillator panels detect a signal. We further reject events where any of the PMT waveforms present large fluctuations in the pedestal region [4].

The time of a PE produced in one of the PMTs of the array, defined using a constant fraction discrimination (CDF) method at a 20% of the pulse height, is calculated relative to the trigger time. For the latter, we use a fixed threshold approach, defined as the time at which the trigger waveform crosses 50% of the height of the SPE pulse, in order to ensure we are computing the time of the first PE. This provides an optimal time resolution for high light levels, as we confirmed experimentally. PMT times are corrected by photon time of flight and cable delays, previously calibrated as described in [4]. Multi-PE contamination on the PMT array is reduced below 5% by selecting hits with a charge between 0.5 PE and 1 PE. The PMT charge is provided by integrating the digitized waveform and the PMT gains are calibrated by characterizing SPE charge distributions, as described in [4].

Using a deionized water target, in order to guarantee a pure Cherenkov emitter, we measured the time resolution to be $(855 \pm 8)$ps FWHM for SPE hits. We validate our MC model by simulating the full setup with the target filled with water and comparing it to our obtained resolution. The prediction underestimates it by 248 ps, so an additional smearing is included in the MC as a correction. This small discrepancy is associated with PMT-to-PMT variations in the transit-time spread and pulse shapes.

The scintillation time profile $p(t)$ is modeled as the sum of two negative exponentials, to represent the decay profile, modulated by an exponential rise:

$$p(t) = R_1 \frac{e^{-t/\tau_1} - e^{-t/\tau_r}}{\tau_1 - \tau_r} + (1 - R_1) \frac{e^{-t/\tau_2} - e^{-t/\tau_r}}{\tau_2 - \tau_r} \quad (6)$$

where $\tau_r$, $\tau_1$ and $\tau_2$ are the rise time, short and long decay times, respectively and $R_1$ is the fraction of the short time component over the total. This model is commonly used in similar studies [9,10] and it has been shown to fit the data well. The parameters are independently determined for each WbLS mixture by implementing this time profile model into the MC and fitting the data.

The detected time profile is fit to the MC prediction using a binned maximum likelihood method. We minimize $-2\log\mathscr{L}$ by iteratively scanning the previously defined parameters. After the minimum has been found, the $1\sigma$ uncertainties are computed by fitting a three-degree polynomial to the individual 1-dimensional scans. This uncertainty, $\sigma_i^*$, on parameter $i$ does not include the correlations, $\rho_{ij}$, with other parameters, $j$. Given that scanning the multi-dimensional likelihood space using MC simulations is extremely compute-intensive, $\rho_{ij}$ are calculated off-line using the analytical form of the scintillation time profile model Eq. (6). The total uncertainty, $\sigma_i$, on a parameter $i$, is given by [1]:

$$\sigma_i = \sigma_i^* \times \prod_j (1 - \rho_{ij}^2)^{-1/2}. \quad (7)$$

This procedure is validated using 100 toy MCs and the final scanned values are bias-corrected. The size of the bias is $-0.28\sigma$ for $\tau_1$, $-0.09\sigma$ for $\tau_2$, $-0.03\sigma$ for $R_1$ and $-0.9\sigma$ for $\tau_r$. The large bias in $\tau_r$ is due to its strong correlation with the normalization of the Cherenkov component, which is already reflected in the poor constraints on the measured values. According to this study, the uncertainties on the different parameters are overestimated by 40% for $\tau_r$, by 32% for $\tau_1$, by 47% for $\tau_2$ and by only 2% for $R_1$. In order to be conservative, the obtained uncertainties from the fit are not reduced by these factors.

**Table 2** Time profile best fit parameters. From top to bottom: rise time, short decay time, long decay time, and fraction of the short decay time component over the total. The quoted uncertainties include the statistical and systematic uncertainties

| WbLS | 1% | 5% | 10% |
| --- | --- | --- | --- |
| $\tau_r$ [ns] | $0.00 \pm 0.06$ | $0.06 \pm 0.11$ | $0.13 \pm 0.12$ |
| $\tau_1$ [ns] | $2.25 \pm 0.15$ | $2.35 \pm 0.13$ | $2.70 \pm 0.16$ |
| $\tau_2$ [ns] | $15.10 \pm 7.47$ | $23.21 \pm 3.28$ | $27.05 \pm 4.20$ |
| $R_1$ | $0.96 \pm 0.01$ | $0.94 \pm 0.01$ | $0.94 \pm 0.01$ |

*3.2.3 Results*

The fit results for the three different WbLS mixtures are shown in Table 2. The time profile distributions for data and MC, using the best fit parameters, are shown in Fig. 4.

The considered sources of systematic uncertainty are the hit-time resolution, the uncertainty in the Cherenkov component, the WbLS light yield and the PPO concentration. The smearing added to the MC due to the underestimation of the time resolution discussed in Sect. 3.2.2 is included in this measurement. Since the size of the Cherenkov component depends on the refractive index of the medium and this has not been measured as a function of the wavelength, we float an extra normalization parameter for the Cherenkov component and treat it as a nuisance parameter. The WbLS light yield could potentially impact the amount of multi-PE hits we observe, biasing our measurement. We compared the MC time distributions obtained for light yield parameters modified by $1\sigma$, as measured in Sect. 3.1. No significant variations were observed, confirming that this effect is negligible in our setup. Potential uncertainties in the reemission probability can affect the shape of the time profile, given that absorbed and reemitted photons get delayed. The reemission probability is a function of the effective concentration of PPO that contributes to the scintillator mechanism, which needs further study [10]. Given that this uncertainty could potentially impact the measured time profile, we compared MC time distributions modeling several high PPO concentrations, demonstrating a negligible impact.

A complementary measurement using X-rays and assuming a three decay exponential model is provided in [12]. Although the three-decay exponential model is favored in [12], we are not sensitive to long decay constants, and hence our choice of a two-decay exponential model. This is ultimately justified by the fact that our measured time profiles are compatible within $1\sigma$ with those extracted in [12] for the prompt 5 ns window, which is the relevant region for Cherenkov and scintillation separation. Beyond that region, our model and [12] diverge both due to the lack of sensitivity to the third exponential decay, which is driven by the large statistical uncertainties in the tail of the distributions, and our MC-based fitting approach, which limits the size of the parameter space (adding a third decay exponential would add an additional dimension to be scanned, complicating the fit procedure to an extent that was deemed unnecessary given our limited sensitivity to the long tail). For this analysis, our two-decay exponential model is used.

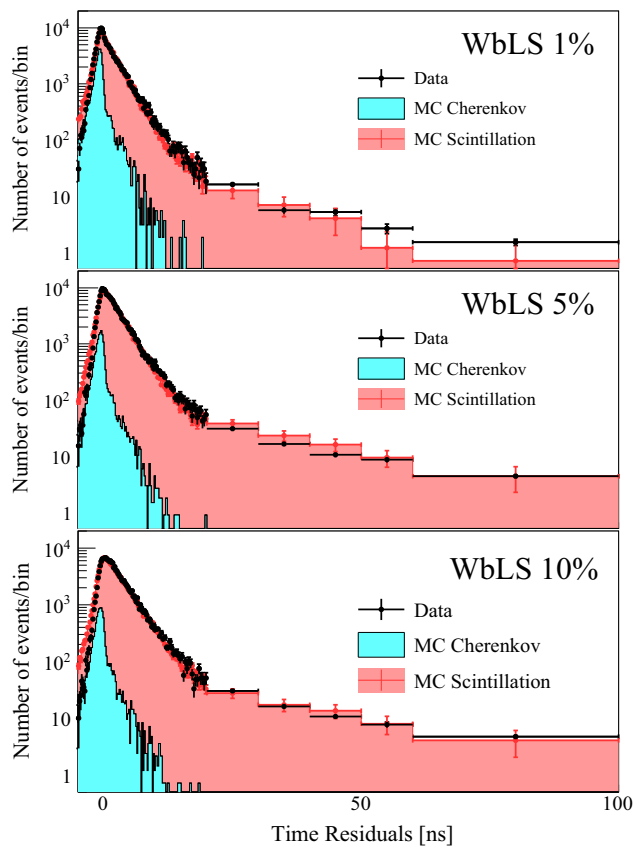

**Fig. 4** Scintillation time profile as detected by the PMT array. Black dots represent data with statistical uncertainties, and colored histograms correspond to MC, which is divided into Cherenkov (blue) and scintillation components (red). MC includes only statistical uncertainties

## 4 Cherenkov and scintillation separation with cosmic muons

The CHESS experiment [4,9] is used in order to quantify the separation between Cherenkov and scintillation light. Using vertical, downward-going cosmic muons we detect the Cherenkov and scintillation light produced in a target filled with WbLS. The strong directional nature of the Cherenkov light projects ring-like structures in the upward-facing PMT array. By using the PMT hit-time information and the number of prompt PEs, we are able to separate the populations of PMTs that detect Cherenkov light from those that do not, thus imaging clear Cherenkov ring structures. This analysis also serves as a validation of our WbLS model in a different configuration, using an independent source in a different energy regime (cosmic muons).

### 4.1 Description of the apparatus

To the geometry described in Sect. 3.2.1, we add two cosmic tags, one above the vessel and another below the PMT array, aligned vertically with the center of the vessel to provide the trigger for through-going vertical cosmic muons (see Fig. 5). The original trigger PMT is removed from the side of the vessel to avoid possible reflections from its front face. The geometry of the setup is optimized so that a cosmic muon passing vertically downwards through the vessel projects a distinct Cherenkov ring in the middle PMTs while it creates isotropic scintillation light detected by the entire PMT array. The data acquisition is triggered off the bottom muon tag and a triple coincidence between the two muon tags and the scintillator panel below the vessel is required offline. This setup is the same as that used in previous studies and further details can be found in [4,9].

### 4.2 Calibrations and analysis method

After the triple coincidence is required, we veto non single-muon events (showers, secondary particles in surrounding materials, etc.) by requiring that there is no signal in any of the other scintillator panels. Due to production of secondary particles that travel through the acrylic block, some residual light is observed in some events in the PMTs closer to the vertical. This effect is reduced by rejecting events whose total number of detected PEs in those PMTs is abnormally large. Events with unstable pedestals are also rejected. More detail on event selection criteria can be found in [4]. After the event selection is applied, 179 events are selected for WbLS 1%, 158 for WbLS 5% and 177 for WbLS 10%, for a 5-week exposure per sample.

In order to compare to our MC prediction, the detection efficiency was calibrated for each PMT. The light collection of each individual PMT depends on the PMT CE and the qual-ity of the optical coupling. The individual photon detection efficiency is calibrated for each PMT of the array by using the scintillation light produced by cosmic muons in a LABPPO 2g/L sample, whose absolute light yield is well known – as measured by SNO+ [15] and confirmed here (Sect. 3.1.3). The average number of PEs per PMT is measured and compared to that predicted by the MC model. A photon detection efficiency parameter is defined for each PMT as the ratio of the average number of PEs in data over that of MC. An independent factor to calibrate the amount of Cherenkov light detected in the PMT array is introduced to take into account systematic uncertainties on the exact ring position and discrepancies of the refractive index with respect to the estimation. A single factor that weights the amount of detected Cherenkov light is included in our MC model. This factor is calibrated using the Cherenkov light produced by cosmic muons in a water target. The ratio of the average number of PEs between data and MC provides the Cherenkov detection efficiency factor, which corresponds to $0.59 \pm 0.04$. This calibration is performed after the individual PMT detection efficiencies have been measured. None of these calibration factors have a significant effect on our WbLS characterization measurements given that we estimate the light yield with a completely independent setup Sect. 3.1 and the time profile measurements are not affected by multi-PE effects nor by changes in the Cherenkov component, due to the fact that the Cherenkov normalization is freely floated during the parameter scans. Furthermore, it was confirmed through simulation that the impact of these parameters on the time distributions is negligible.

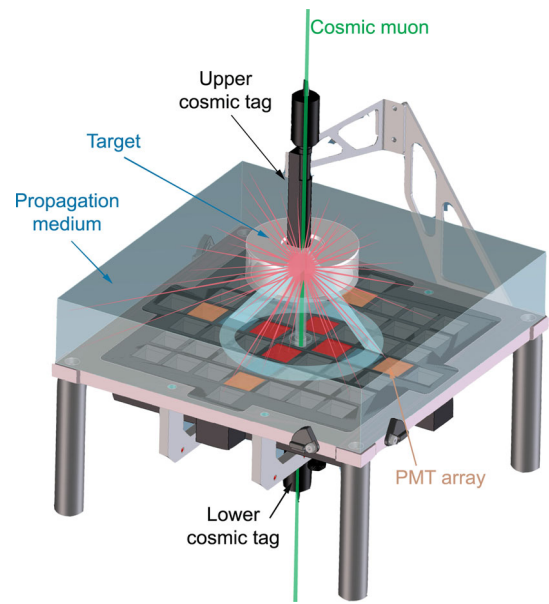

**Fig. 5** Configuration for the Cherenkov and scintillation separation study with cosmic muons. The detected Cherenkov rings are expected to lay on the middle PMTs, highlighted in blue. See Sect. 4.1 for a more detailed description. In the same fashion than Fig. 1, the scintillator panels are not drawn and their layout can be seen in [4]

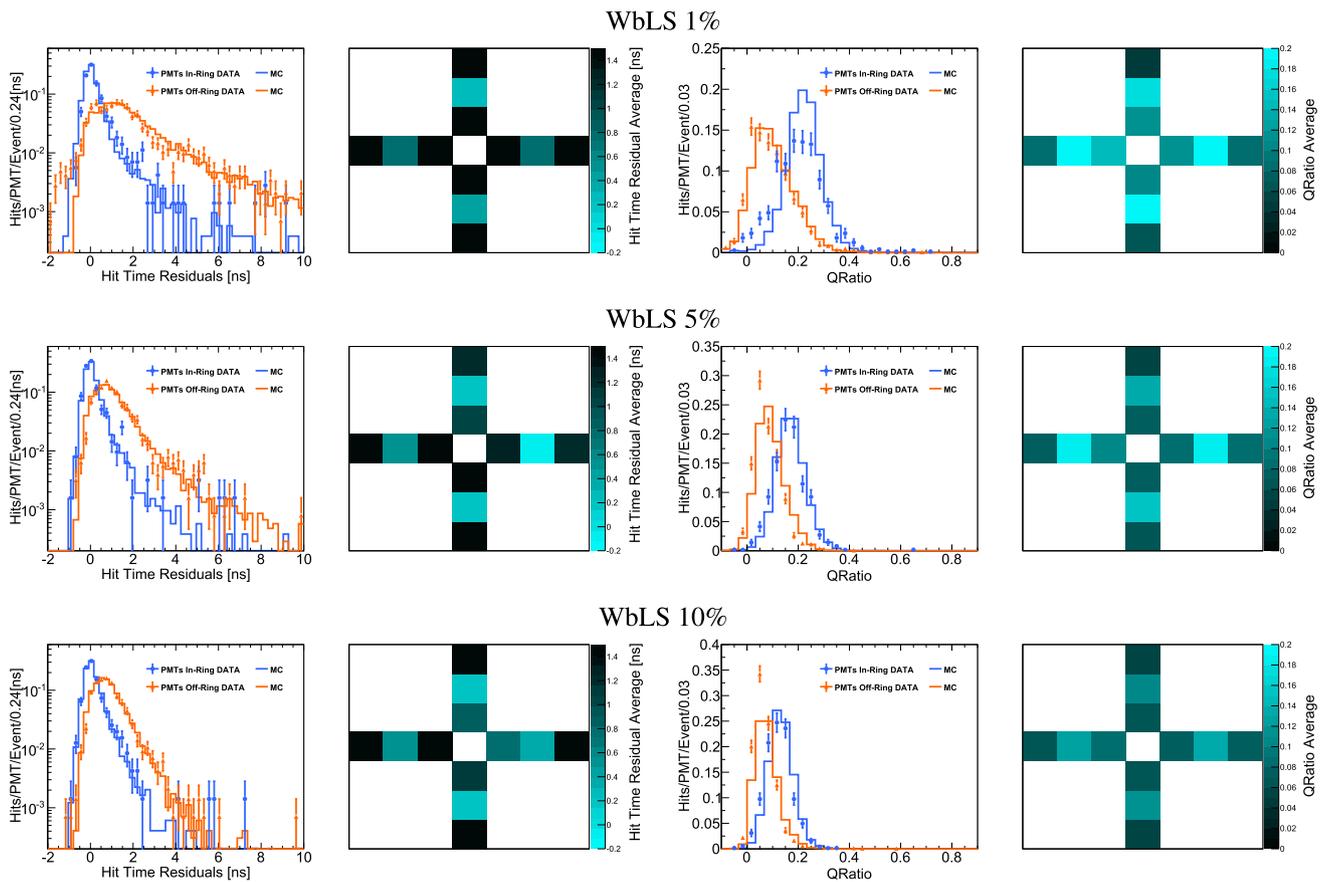

**Fig. 6** Cherenkov/scintillation separation in hit-time residuals and charge ratio for (top) 1%, (middle) 5% and (bottom) 10% WbLS. (Leftmost): PMT hit-time residual distributions for PMTs in the Cherenkov ring and outside it; (left-middle): PMT hit-time residuals vs PMT position, averaged across the data set. (Right-middle): $Q_r$ (ratio of the charge integrated in a 5 ns window over the total charge) distributions for PMTs in the Cherenkov ring and outside it; and (right-most) $Q_r$ vs PMT position averaged across the data set

We evaluate the Cherenkov and scintillation separation as a function of two observables: the PE hit-time and the fraction of prompt light.

The time of the first PE in each PMT hit is calculated using a fixed threshold of 25% of the SPE pulse height, as measured during our calibration campaign for the individual PMTs. The time residuals are defined as the hit time relative to the event time, calculated as the average of the first 4 earliest PMT hit times. The hit times are corrected by cable delays and time of flight. A measurement of our time resolution is performed by using cosmic muons in a water target, given that the fast nature of the Cherenkov emission provides a sharp light pulse with a width under tens of picoseconds. The measured time resolution is $(275 \pm 2)$ ps for data and $(262 \pm 2)$ ps for MC. (This is improved over that reported in Sect. 3.2.2 due to the multi-PE nature of these events). In order to account for the slight discrepancy, an 82 ps Gaussian smearing is applied to the MC time distributions.

The fraction of early light collected in a PMT, $Q_r$, is defined as the ratio of charge integrated in a prompt 5 ns window around the event time to the total integrated charge [4]. PMT hits due to pure Cherenkov photons would have a larger $Q_r$ value than those due to slower scintillation light. Given that the SPE PMT pulses for the CHESS PMTs are tens of nanoseconds wide, the value of $Q_r$ even for prompt photons is smaller than unity, and the exact value depends on the precise shape of the PMT waveform. We model PMT pulses using an analytical form that describes the average behaviour for all the electronic channels [4], but small deviations in the pulse shape with respect to this form can affect the predicted values of $Q_r$. We quantify this effect by calibrating $Q_r$ using a dataset of cosmic muons in water, which provides a source of prompt Cherenkov light. A correction factor is calculated as the ratio of the average $Q_r$ in data over that in MC, which yields a factor of 1.34. This is applied as a correction to the MC for the WbLS samples.

### 4.3 Results

The hit-time residual distributions are shown in Fig. 6 for the three WbLS concentrations, broken down by PMT radial position. We expect the Cherenkov ring to fall on the middle

**Table 3** Results of Cherenkov and scintillation separation for cosmic muons and the three different versions of WbLS for the time residual (top) and $Q_r$ (bottom) methods. Values for the Cherenkov detection efficiency $\varepsilon_C$ and the scintillation contamination $f_s$ are shown for a common cut value (reference cut) and for the optimal cut that maximizes the figure of merit (Eq. (8))

| Time | | | |
|---|---|---|---|
| LS% | 1% | 5% | 10% |
| Reference cut | 0.2 ns | | |
| $\varepsilon_C$ | 75 ± 4% | 84 ± 4% | 80 ± 4% |
| $f_s$ | 22 ± 2% | 19 ± 2% | 25 ± 2% |
| Optimal cut | 0.8ns | 0.2ns | 0.2ns |
| $\varepsilon_C$ | 88 ± 4% | 84 ± 4% | 80 ± 4% |
| $f_s$ | 27 ± 2% | 19 ± 2% | 25 ± 2% |
| $Q_r$ | | | |
| LS% | 1% | 5% | 10% |
| Reference cut | 0.1 | | |
| $\varepsilon_C$ | 86 ± 4% | 85 ± 4% | 65 ± 4% |
| $f_s$ | 30 ± 7% | 26 ± 8% | 22 ± 10% |
| Optimal cut | 0.1 | 0.1 | 0.05 |
| $\varepsilon_C$ | 86 ± 4% | 85 ± 4% | 96 ± 5% |
| $f_s$ | 30 ± 7% | 26 ± 8% | 45 ± 5% |

PMTs (in-ring), while the rest would detect only scintillation (off-ring). The in-ring PMTs show a hit-time residual distribution shifted towards earlier times with respect to the off-ring distribution, as expected given the faster nature of Cherenkov light. The average of the hit-time residuals are also shown in Fig. 6 for each PMT position, displaying clear ring structures.

$Q_r$ distributions are presented in Fig. 6, which also show a clear separation between the two PMT populations. By plotting $Q_r$ values averaged over events, Cherenkov ring structures can be identified (Fig. 6). Residual disagreement between data and MC is attributed to individual PMT variations in the exact pulse shape, as discussed in Sect. 4.2. Since the MC is not used in the evaluation of the separation of Cherenkov and scintillation signals, these discrepancies do not affect the results quoted in this paper. For future detectors, this highlights the importance of a detailed understanding of PMT pulse shape in such analyses.

Given a time residual value $t$, we define the Cherenkov detection efficiency $\varepsilon_C(t)$ as the fraction of in-ring PMT hits that occur before that time, $t$, such that it represents the fraction of detected Cherenkov hits. The scintillation contamination $f_s(t)$ within the prompt time window is defined as the fraction of hits occurring prior to $t$ that are observed on the off-ring PMTs, such that it represents the scintillation contribution to this selection. We identify a time residual cut $t^c$ that maximizes $\varepsilon_C(t)$ while reducing the scintillation fraction $f_s(t)$, by maximizing the figure of merit:

$$\varepsilon_C \times (1 - f_s). \tag{8}$$

The same technique is applied to $Q_r$. The Cherenkov detection efficiency $\varepsilon_C(Q_r)$ is now defined as the fraction of in-ring PMT hits that fall above $Q_r$, and the scintillation fraction $f_s(Q_r)$ is defined as the fraction of hits above $Q_r$ that are on off-ring PMTs. An optimal value $Q_r^c$ is found by maximizing the figure of merit.

Table 3 shows the $\varepsilon_C$ and $f_s$ values for reference cut values of t and $Q_r$ and for the optimal cut values. A high Cherenkov detection efficiency with a reasonably low scintillation light fraction is achieved. The non-linear behavior with concentration is due to the slightly different time profile measured for the different mixtures. In particular, we observe that the WbLS 10% time constants are longer than the other WbLS concentrations, which makes the Cherenkov and scintillation separation perform similarly to the other cases despite the higher light yield.

When compared to pure LAB and LABPPO 2g/L [9], an overall improvement on the Cherenkov and scintillation separation is achieved. The Cherenkov detection efficiency for WbLS is larger than that of pure LABPPO 2g/L (70% and 63% for time and $Q_r$ respectively), and at the level of that of pure LAB (83% and 96% for time and $Q_r$ respectively). The scintillation fraction is smaller than that of pure LABPPO 2g/L (36% and 38% for time and $Q_r$ respectively), demonstrating the potential to select a more pure sample of Cherenkov hits using separation in time and prompt charge.

## 5 Conclusion and discussion of the results

We have measured the absolute scintillation light yield and emission time profile of three mixtures of WbLS at different concentrations of LS in water. Moreover, we have demonstrated improved performance for separation of Cherenkov and scintillation signals with respect to that achieved in LABPPO 2g/L for cosmic muons in a small-scale target (centimeters). This publication is also followed by the release of the first data-driven MC model of the different characterized WbLS mixtures, compatible with the software RAT-PAC [14] and intended for public use.

The observables measured by this work comprise several of the critical factors for determining separation of Cherenkov and scintillation signals, and detector performance of liquid scintillator detectors at large scales (meters to 10 s of meters). Other critical parameters for predicting performance at large scales are the absorption and scattering lengths, and dispersion (as determined by the refractive index). Precise evaluation of WbLS performance requires a complete understanding of these effects and extrapolation using a detailed MC model including detector size and geometry. This complex study is out of scope of the current work. The effort is ongoing, and a separate publication will be prepared. A qualitative discussion of large-scale performance in light of our results is included below.

When compared to LABPPO 2g/L, a standard liquid scintillator used in detectors like SNO+ [22], two points must be remarked. First, the measured light yield is smaller than that of pure scintillator, which eases the Cherenkov and scintillation separation but reduces the light production and, thus, the energy resolution. Nevertheless, the light yield is larger than expected from a naive scaling of the scintillator content, by a factor between 1.5 and 2.5, which is an advantage for physics topics for which energy resolution is important. Second, the extracted WbLS time profiles indicate a faster photon emission rise time and first decay component with respect to LABPPO 2g/L. A faster scintillation component is beneficial for vertex resolution, but induces a greater overlap with the prompt Cherenkov emission peak, resulting in a less efficient separation. These two points refer to competing effects that need to be carefully evaluated for large-scale detector scenarios. Given the flexible nature of the composition of WbLS, a mixture to ensure an optimal separation and scintillation light yield for large detectors will be pursued.

**Acknowledgements** This material is based upon work supported by the U.S. Department of Energy, Office of Science, Office of High Energy Physics, under Award Number DE-SC0018974. Work conducted at Lawrence Berkeley National Laboratory was performed under the auspices of the U.S. Department of Energy under Contract DE-AC02-05CH11231. The work conducted at Brookhaven National Laboratory was supported by the U.S. Department of Energy under contract DE-AC02-98CH10886. The project was funded by the U.S. Department of Energy, National Nuclear Security Administration, Office of Defense Nuclear Nonproliferation Research and Development (DNN R&D). The authors would like to thank the SNO+ collaboration for providing data on the optical properties of LAB/PPO, including the light yield, absorption and reemission spectra, and refractive index. Special thanks to Dr. Zara Bagdasarian for her thorough review and useful feedback on the optical model section.

**Data Availability Statement** This manuscript has no associated data or the data will not be deposited. [Authors' comment: The data obtained in this experiment is only relevant in the context of our prototype and will not be publicly available. Results will be conveniently deposited in a public repository [14].]